\documentclass[aps,nofootinbib, superscriptaddress, showpacs,showkeys,floatfix]{revtex4}
\usepackage{epsfig,graphicx}
\usepackage{graphicx}

\begin{document}
\title{The dynamics of the early universe and the initial conditions for inflation in a model with radiation and a Chaplygin gas}

\author{G. A. Monerat\footnote{E-mail: monerat@uerj.br}} 
\author{G. Oliveira-Neto\footnote{E-mail: gilneto@fat.uerj.br}}
\author{E. V. Corr\^{e}a Silva\footnote{E-mail: evasquez@uerj.br}} 
\address{Departamento de Matem\'{a}tica e Computa\c{c}\~{a}o, 
Faculdade de Tecnologia, \\ 
Universidade do Estado do Rio de Janeiro, Estrada Resende-Riachuelo, s/n$^o$, Morada da Colina \\
CEP 27523-000, Resende-RJ, Brazil.}

\author{L. G. Ferreira Filho\footnote{E-mail: gonzaga@fat.uerj.br}}
\address{Departamento de Mec\^{a}nica e Energia, 
Faculdade de Tecnologia,\\ 
Universidade do Estado do Rio de Janeiro, Estrada Resende-Riachuelo, s/n$^o$, Morada da Colina \\
CEP 27523-000 , Resende-RJ, Brazil.}

\author{P. Romildo Jr.\footnote{E-mail: pauloromildo@yahoo.com.br}}
\address{Departamento de F\'{\i}sica, Instituto de Ci\^{e}ncias Exatas, 
Universidade Federal de Juiz de Fora, \\
CEP 36036-330, Juiz de Fora, Minas Gerais, Brazil.}

\author{J. C. Fabris\footnote{E-mail: fabris@cce.ufes.br}} 
\author{R. Fracalossi\footnote{E-mail: rfracalossi@cce.ufes.br}}
\author{S. V. B. Gon\c{c}alves\footnote{E-mail: sergio@cce.ufes.br}}
\address{Departamento de F\'{\i}sica, Centro de Ci\^{e}ncias Exatas,\\ 
Universidade Federal do Esp\'{\i}rito Santo, Avenida Fernando Ferrari s/n
Goiabeiras \\
CEP 229060900 - Vit\'{o}ria, ES - Brazil.}

\author{F. G. Alvarenga\footnote{E-mail: flavioalvarenga@ceunes.ufes.br}}
\address{Departamento de Engenharia e Ci\^{e}ncias Exatas (DECE),\\
Centro Universit\'{a}rio Norte do Esp\'{\i}rito Santo,\\ 
Universidade Federal do Esp\'{\i}rito Santo,\\ 
Rua Humberto de Almeida Franklin, 257, Bairro Universitário\\
CEP 29933-480, São Mateus, ES - Brazil.}

\date{\today}

\begin{abstract}
{\footnotesize The modeling of the early universe is done through the quantization of a
Friedmann-Robertson-Walker model with positive curvature. The material content consists of two
fluids: radiation and Chaplygin gas. The quantization of these models is made by following the
Wheeler and DeWitt's prescriptions. Using the Schutz formalism, the time notion is recovered and
the Wheeler-DeWitt equation transforms into a time dependent Schr\"{o}dinger equation, which
rules the dynamics of the early universe, under the action of an effective potential $V_{ef}$.
Using a finite differences method and the Crank-Nicholson scheme, in a code implemented in the
program OCTAVE, we solve the corresponding time dependent Schr\"{o}dinger equation and
obtain the time evolution of a initial wave packet. This wave packet satisfies appropriate
boundary conditions. The calculation of the tunneling probabilities shows that the universe may
emerge from the Planck era to an inflationary phase. It also shows that, the tunneling
probability is a function of the mean energy of the initial wave packet and of two parameters
related to the Chaplygin gas. We also show a comparison between these results and those
obtained by the WKB approximation.}
\end{abstract}



\maketitle
\newpage
\section{Introduction}

One of the main purposes of quantum cosmology is to fix the initial conditions that will determine the later behavior of the Universe \cite{grishchuk}.
In quantum cosmology, the Universe is treated as a quantum system. Since the cosmological
scenario is based on general relativity, the dynamical variables are directly related to the
geometry of the space-time. This implies a functional equation, the Wheeler-DeWitt equation,
on all possible geometries \cite{wheeler1}. Such functional equation has in general no exact
solution. A possible way to circumvent this technical limitation is to freeze out an infinite
number of degrees of freedom, remaining at the end with what is called a minisuperspace, a
configuration space with few degrees of freedom, which may admit exact solutions. Even if such
procedure may be seen as a drastic limitation of the original problem, it allows at least to
extract some precise predictions concerning the evolution of the early Universe. One of the
hopes in such a program is to obtain initial conditions for the classical evolution leading to
an inflationary expansion of the Universe.
\par
Yet, the limitation to a minisuperspace approach does not eliminate other difficulties in quantum cosmology. The Wheeler-DeWitt equation has no explicit time variable \cite{isham}. This
is a consequence of the fact that the Einstein-Hilbert action represents a constrained system
which is invariant by time reparametrization. Moreover, the limitation to a few number of
degrees of freedom, the restriction to the minisuperspace, does not assure that the final
equation will admit closed solutions. Frequently, it is necessary to look for approximative
solutions \cite{monerat6} using, for example, perturbative methods. The usual WKB method of
ordinary quantum mechanics is one of the possible methods to extract predictions of a quantum
cosmological system  \cite{colistete}. It is a quite attractive procedure since it allows to
introduce a time variable which is, otherwise, absent.
\par
In some specific situations it is possible to recover the notion of a time variable in the minisuperspace approach. One example is when gravity is coupled to a perfect fluid, which is
expressed through the Schutz's variable \cite{schutz}: the conjugate momentum associated with
the fluid's variable appears linearly in the Lagrangian, which allows to recover a genuine
Schr\"odinger equation \cite{fluido}. Since now the quantum system has an explicit dynamics, it
is possible to study the emergence of the classical Universe in a primordial era. An important
question can be addressed in this way: what are the conditions to have an initial inflationary
phase in the evolution of the Universe?
\par
We will study this problem in this paper. Gravity will be coupled to a radiative perfect fluid,
whose dynamic variables will be connected with the time evolution of the quantum system. Aside
the radiative fluid, the Chaplygin gas will be included in this scenario. The Chaplygin gas
is an exotic fluid exhibiting negative pressure which depends on the inverse of the density
\cite{chaplygin}. The interest on the Chaplygin gas fluid has been, until now, mainly connected
with the existence of a dark energy component in the Universe today \cite{fabris}. In this
sense, it leads to a scenario competitive with the $\Lambda CDM$ model. However, it can be of
interest for the early Universe, and this due to a main reason: the Chaplygin gas can be
obtained from the string Nambu-Goto action re-expressed in the light cone coordinate
\cite{jackiw}. Since string theory manifests, in principle, its main features in very high
energy levels, as it occurs in the early Universe, exactly when the quantum effects must be
relevant, a quantum cosmological study of the Chaplygin gas is of course relevant. Moreover,
the Chaplygin gas admits a de Sitter asymptotically phase, that is, it tends to a cosmological
constant, being in one sense a much richer structure. In that sense, one may consider the
present paper a generalization of a previous paper made by some of us, where instead of the
Chaplygin gas we used a positive cosmological constant \cite{gil}. For all these reasons, a
quantum cosmological scenario using radiative fluid and the Chaplygin gas is a very attractive
configuration by its own.
\par
A quantum cosmological model with two fluids, as it will be treated here, implies considerable
technical difficulties. In particular, no analytical solution can be expected. Hence, we are
obliged to consider a perturbative analysis, like WKB \cite{colistete}, or a numerical
integration of the equations. We will concentrate, here, in the latter possibility. It will be
used a finite differences method and the Crank-Nicholson scheme, in a code implemented in the
program OCTAVE. Using such program, we will compute the tunneling probability (TP) from ability (TP) from a
quantum phase to the classical regime, the potential barrier being represented by the density
of the Chaplygin gas in terms of the scale factor. When the system emerges from the potential barrier, it finds itself in an inflationary phase. The main results of our paper, besides the
solution of the present quantum cosmological model, is related to the determination of the 
initial conditions for the classical evolution of the universe. Here, for the present model, we would gain information on what is the most probable amount of radiation in the initial evolution of the classical universe and the most probable values of the parameters of the Chaplygin Gas. If we take in account that one of those parameters tends to the cosmological constant ($\Lambda$), for great values of the scale factor, we would gain information on the most probable value of $\Lambda$. Another important result, derived here, is the fixation of
the initial value of the scale factor and its conjugated momentum leading to a classical
inflationary evolution of the scale factor.

In fact, the quantum cosmology of a closed FRW model coupled to a Chaplygin gas have already been studied in the literature \cite{moniz}. We may mention several differences between our
work and Reference \cite{moniz}. In their work, the authors considered the generalized Chaplygin gas as the only source of matter. Here, we consider the standard Chaplygin gas and radiation as our sources of matter. We use the Schutz formalism in order to introduce a time
variable and transform the Wheeler-DeWitt equation in a Schr\"{o}dinger equation. Most importantly, we solve the Wheeler-DeWitt equation exactly using a numerical procedure, what was even suggested by the authors of Reference \cite{moniz}. On the other hand, the authors of \cite{moniz} use an approximated potential to describe the Chaplygin gas and solve the
resulting Wheeler-DeWitt equation, to that approximated potential, in the WKB approximation. 
In despite of all those differences between the models and the treatments used here and in \cite{moniz}, it is important to mention that we have obtained few qualitative agreements concerning the tunneling probability as a function of the parameters $A$ and $B$. Their approximated, analytical expression for $TP$ as a function of $A$, or the cosmological constant, grows with $A$ for the boundary conditions of the tunneling wave-function \cite{vilenkin}. Similarly, their approximated, analytical expression for $TP$ as a function 
of $B$, grows with $B$ for the boundary conditions of the tunneling wave-function \cite{vilenkin}. Those results are in agreement with our numerical curves for $TP$ as a function of $A$ and as a function of $B$.

\par
This paper is organized as follows. In next section, the classical model is presented. In section 3, the classical model is quantized. In section 4, the wave packet describing the quantum system is constructed. In section 4 and 5, the tunneling effect is investigated and the initial conditions for inflation are settled out. The conclusions are presented in section 6.

\section{The Classical Dynamics}

\begin{figure}
\includegraphics[width=6cm,height=10cm, angle=-90]{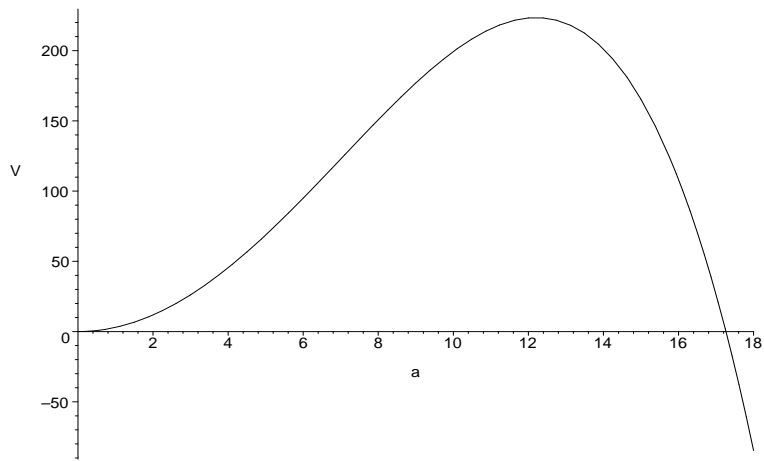}
\vspace{0.8cm}
\caption{Behavior of the effective potential $V_{ef}(a)$ for $A=0.001$ and $B=0.001$.}
\label{potential}
%
%
\end{figure}

The homogeneous and isotropic Friedmann-Robertson-Walker (FRW) models with positive curvature
of the spatial sections, radiation and Chaplygin gas, may be represented by a Hamiltonian with
two degrees of freedom in the form
\begin{equation}
H = \frac{1}{12}p_{a}^2+V_{ef}(a)-p_{T},
\label{sec1eq1}
\end{equation}

\noindent
where $p_{a}$ and $p_{T}$ are, respectively, the moments canonically conjugated to the scale factor
and to the variable that describes the perfect fluid, and $V_{ef}(a)$ is an effective potential, 
Figure \ref{potential}).  We are working in the unit system where $G=1$, $c=1$
and $\hbar=1$. $V_{ef}(a)$ is given by,
\begin{equation}
V_{ef}(a) = 3a^2-\frac{a^4}{\pi}\sqrt{A+\frac{B}{a^6}},
\label{sec1eq2}
\end{equation}

\noindent
where $A$ and $B$ are parameters associated to the Chaplygin gas. The Hamilton equations,
\begin{equation}
\left\{
\begin{array}{lll}
\dot{a} &=& \frac{\displaystyle\partial H}{\displaystyle\partial p_{a}} = \frac{1}{6}p_{a}\, ,\\ \nonumber
& & \\
\dot{p_{a}} &=& -\frac{\displaystyle\partial H}{\displaystyle\partial a} = - 6a
+ \left(\frac{\displaystyle 4a^3}{\pi}\right)\sqrt{\displaystyle A+\frac{B}{a^6}} - \frac{\displaystyle 3B}{\displaystyle \pi\ a^3\, \sqrt{A+\frac{B}{a^6}}}\, ,
\\ \nonumber
& & \\
\dot{T} &=& \frac{\displaystyle\partial H}{\displaystyle\partial p_{T}} = -1\, ,\\ \nonumber
& & \\
\dot{p_{T}} &=& -\frac{\displaystyle\partial H}{\displaystyle\partial T} =0\, ,\\
\end{array}
\right.
\label{sec1eq3}
\end{equation}

\noindent
rule the classical dynamics of these models and are completely equivalent to the Einstein equations. The dot stands for ordinary differentiation relative to the conformal time $\tau$. Combining the first two equations in (\ref{sec1eq3}) we have,
\begin{equation}
\ddot{a} = 2\frac{\displaystyle a^3}{\displaystyle 3\pi}\sqrt{\displaystyle A+\frac{B}{a^6}}-\left[
a+\frac{1}{\displaystyle 2\pi a^3\sqrt{\displaystyle A+\frac{B}{a^6}}}\right].
\label{sec1eq7}
\end{equation}

According to (\ref{sec1eq7}), the universe has an accelerated phase ($\ddot{a}>0$) for big values of the scale factor ($a^6 >> B/A$) and a de-accelerated one ($\ddot{a}<0$) when  $a^6 << B/A$. The solutions of (\ref{sec1eq7}) for the latter phase are described by trigonometric functions.

\noindent

\section{The quantization of the model}

The Planck era will be described by the quantization in the minisuperspace of that model, following the prescription proposed by Wheeler and De Witt \cite{wheeler1}. This description is based on the wave function of the universe, which depends on the canonical variables $\hat{a}$ e $\hat{T}$,
\begin{equation}  
\Psi\, =\, \Psi(\hat{a} ,\hat{T} )\, .
\label{sec3eq1}
\end{equation}

The quantization process begins by promoting the canonical moments $p_{a}$ and $p_{T}$ to operators and imposing that the Hamiltonian annihilates the wave function,
\begin{equation}
p_{a}\rightarrow -i\frac{\partial}{\partial a}\hspace{0.2cm},\hspace{0.2cm} 
\hspace{0.2cm}p_{T}\rightarrow -i\frac{\partial}{\partial T}\hspace{0.2cm}, \hspace{0.2cm}\hspace{0.2cm}\hat{H}\Psi(a,\,T)=0.
\label{sec3eq2}
\end{equation}

Then, the quantum dynamics turns out to be ruled by the Wheeler-DeWitt equation, which in this case, assumes the form of a time dependent Schr\"{o}dinger equation,
\begin{equation}
\bigg(\frac{1}{12}\frac{{\partial}^2}{\partial a^2} - 3a^2 + \frac{a^4}{\pi}\sqrt{A+\frac{B}{a^6}}\bigg)\Psi(a,\tau) = -i \, \frac{\partial}{\partial \tau}\Psi(a,\tau),
\label{sec3eq3}
\end{equation}

\noindent
where we impose the reparametrization $\tau= -T$.

The operator $\hat{H}$ is self-adjoint \cite{lemos} in relation to the internal product, 

\begin{equation}
(\Psi ,\Phi ) = \int_0^{\infty} da\, \,\Psi(a,\tau)^*\, \Phi (a,\tau)\, ,
\label{sec3eq4}
\end{equation}
if the wave functions are restricted to the set of those satisfying either 
$\Psi (0,\tau )=0$ or $\Psi^{\prime}(0, \tau)=0$, where the prime $\prime$
means the partial derivative with respect to $a$. Here, we consider wave 
functions satisfying the former type of boundary condition and we also 
demand that they vanish when $a$ goes to $\infty$.

The Wheeler-DeWitt equation (\ref{sec3eq3}) is a Schr\"{o}dinger equation for
a potential with a barrier. We solve it numerically using a finite
difference procedure based on the Crank-Nicholson method \cite{crank}, 
\cite{numericalrecipes} and implemented in the program GNU-OCTAVE \cite{gonzaga}.
Following the discussion in Ref. \cite{gil}, we also use, here, the norm 
conservation as the criterion to evaluate the reliability of our numerical 
calculations of the wave-packet's time evolution. Therefore, we have 
numerically calculated the norm of the wave packet for different times. The 
results thus obtained show that the norm is preserved. Numerically, one can only 
treat the {\it tunneling from something} process, where one gives a initial wave
function with a given mean energy, very concentrated in a region next to 
$a=0$. That initial condition fixes an energy for the radiation and the 
initial region where $a$ may take values. Our choice for the initial wave
function will be described in the following section.

\section{Wave Packets}

As the initial condition, we have chosen the following wave function,
\begin{equation}
\Psi(a,0) =\frac{8 \sqrt[4]{2}\, {E_{m}}^{3/4}\, a\, e^{-4a^2E_m}}{\sqrt[4]{\pi}}\,
\label{sec4eq1}
\end{equation}

\noindent
where $E_{m}$ represents the mean kinetic energy of the wave packet associated to the energy of the radiation fluid. Besides, this initial condition is normalized as follows $\int_{0}^{\infty}|\Psi(a,0)|^2 da=1$.

The portion of the wave function that tunnels the potential barrier, propagates to infinity in the positive scale factor direction, almost like a free particle, as time goes to infinity. However, we must specify a limit, in the scale factor direction, in order to perform the numerical integration of the Schr\"{o}dinger equation. In this case, this value was fixed in $a=30$.  The behavior of these wave packets and their time evolution show that they are well defined in the whole space, even when $a$ goes to zero. 

\begin{figure}[htb]
\includegraphics[width=8cm,height=12cm, angle=-90]{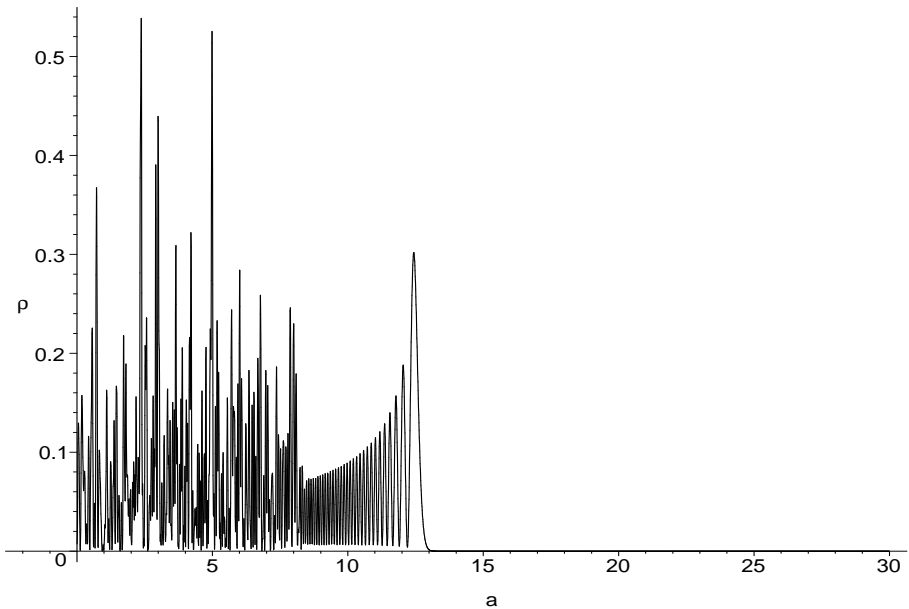}
%
%
\caption{$|\Psi(a,t_{max})|^2 \equiv \rho$, for 
$A=0.001$, $B=0.001$ and $E_m=220$ at the moment $t_{max}$ when $\Psi$ reaches 
infinity, located at $a=30$. }
\label{f1}
\end{figure}

As an example, we fix the values of the system parameters $A=0.001$ and $B=0.001$. We choose
the initial wave packet with mean energy $E_{m}=220$, which gives an initial amplitude of
$5.901650048$. This wave packet will evolve in time, obeying the Schr\"{o}dinger equation
(\ref{sec3eq3}), and reaches the potential barrier at the point $a_{1}=11.41476507$. In this
case, the potential barrier has a maximum value $V_{max}=223.5282023$ at $a_{max}=12.20732711$. The figures \ref{f1} show the time evolution of these wave packets ( $|\Psi(a,0)|^2$), which indicates the possibility of the universe to have a posterior inflationary phase via a tunneling mechanism. With this choice for the parameters, the universe emerges at the right side of the potential barrier with the size $a_2=12.95147904$.

\section{The tunneling effect on the early universe and the initial conditions for inflation}

\subsection{Tunneling probability as a function of $E_m$ and comparison with WKB}
\label{subsec:tpenergy}

In order to evaluate the tunneling probability as a function of the mean energy $E_{m}$, one
needs to fix the values of $A$ and $B$ and repeat the calculations for different
values of $E_{m}$ in $\Psi(a,0)$. For each value of $E_{m}$, the turning point associated to the
potential barrier, to the left ($a_{1}$) and to the right ($a_{2}$) are calculated. 

The probability to find the universe to the right side of the barrier is defined as \cite{gil},

\begin{equation}
\label{sec5eq1}
TP = {\int_{a_{2}}^{\infty} |\Psi(a,tmax)|^2 da \over 
\int_{0}^{\infty} |\Psi(a,tmax)|^2 da} \, ,
\end{equation}
where, as we have mentioned above, numerically $\infty$ has to be fixed to a maximum value of
$a$. Here, we are working with $a_{max} = 30$. It is useful to remember that the denominator of
the equation (\ref{sec5eq1}) is equal to $1$, since the wave function is normalized to unity.

\begin{figure}
\includegraphics[width=8cm,height=12cm, angle=-90]{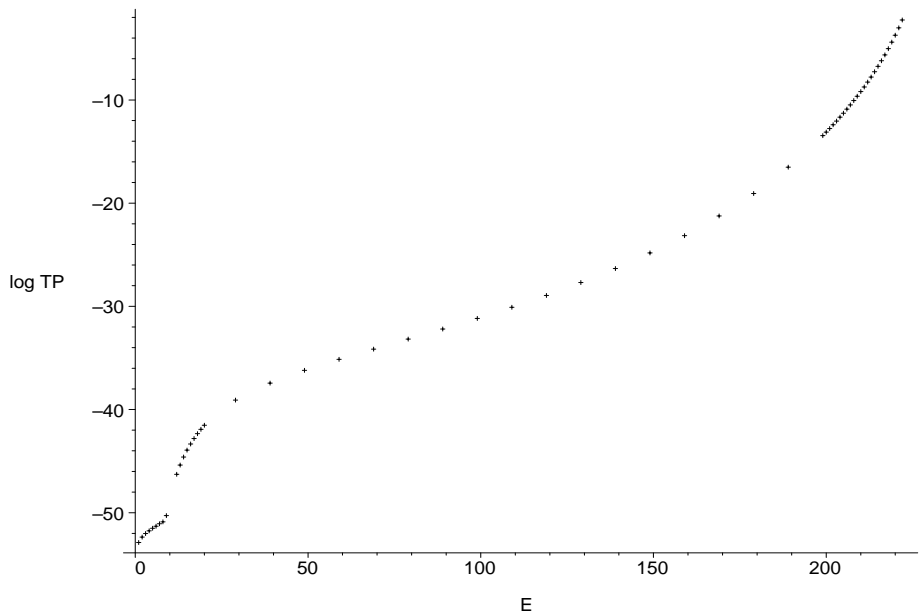}
\vspace{0.8cm}
\caption{Tunneling probability as a function of the mean energy of the initial wave packet for $A=0.001$, $B=0.001$. }
\label{f2}
%
%
%
\end{figure}

In order to evaluate the tunneling probability as a function of the mean energy $E_{m}$, one
needs to fix the values of $A$ and $B$ and repeat the calculations for different
values of $E_{m}$ in $\Psi(a,0)$. For each value of $E_{m}$, the turning point associated to the
potential barrier, to the left ($a_{1}$) and to the right ($a_{2}$) are calculated. 
For all cases, we consider the situation where $E_m$ is smaller than the
maximum value of the potential barrier. From that numerical study we 
conclude that the tunneling probability grows with $E_m$ for fixed $A$ and $B$.
As an example, we consider $60$ values of the radiation energy for fixed $A=0.001$ and
$B=0.001$. For this choice of $A$ and $B$ the potential barrier has its 
maximum value equal to $196.0471982$. In order to study the tunneling process, we fix 
the mean energies of the various $\Psi(a,0)$'s eq. (\ref{sec4eq1}) to be smaller 
than that value. In table \ref{tabela1}, in the appendix, we can see, 
among other quantities, the different values of the energy $E_m$, $TP$, 
$a_1$ and $a_2$ for each energy. In figure \ref{f2}, we see the 
tunneling probability as functions of $E_m$, for this particular example. Due 
to the small values of some $TP'$s, we plot the logarithms of the $TP'$s 
against $E_m$.  

Based on the detailed discussion made in Ref. \cite{gil}, it is not difficult to 
understand under what circumstances the tunneling probability computed with the WKB 
wave-function ($TP_{WKB}$) agrees with $TP$ Eq. (\ref{sec5eq1}). In order to see
it, let us consider that, initially, we have an incident wave ($\Psi_I$) that reaches 
the potential barrier at $a_1$. Then, part of $\Psi_I$ is transmitted to $\infty$ 
($\Psi_T$) and part is reflected ($\Psi_R$). In the present problem, we have an 
infinity potential wall at $a=0$ because the scale factor cannot be smaller than zero. 
It means that $\Psi_R$ cannot go to $-\infty$, as was assumed in order to compute the 
$TP_{WKB}$. Instead, $\Psi_R$ will reach the infinity potential wall at $a=0$ and will 
be entirely reflected back toward the potential barrier, giving rise to a new incident
wave ($(\Psi_R)_I$). The new incident wave $(\Psi_R)_I$ reaches the potential barrier 
at $a_1$ and is divided in two components. A reflected component which moves toward the 
infinity potential wall at $a=0$ ($((\Psi_R)_I)_R$) and a transmitted component which 
moves toward $\infty$ ($((\Psi_R)_I)_T$). $((\Psi_R)_I)_T$ will contribute a new amount
to the already existing $TP$ due to ($\Psi_T$). Therefore, the only way
it makes sense comparing $TP$ with $TP_{WKB}$ is when we let the system
evolve for a period of time ($\Delta t$) during which $\Psi_R$ cannot be 
reflected at $a=0$ and come back to reach the potential barrier. It is clear
by the shape of our potential that the greater the mean energy $E_m$ of the
wave-packet (\ref{sec4eq1}), the greater is ($\Delta t$). For the present situation 
$TP_{WKB}$ is defined as \cite{merzbacher},

\begin{equation}
TP_{WKB} = \frac{4}{\left[\displaystyle 2e^{\displaystyle\int_{a_{1}}^{a_{2}}2\sqrt{3(V-E)}da}+\frac{1}{2}\, \frac{1}{\displaystyle e^{\displaystyle\int_{a_{1}}^{a_{2}}2\sqrt{3(V-E)}da}}\right]^2}.
\label{sec5eq2}
\end{equation}

As an example, in Table \ref{comparison}, we show a comparison between $TP$ 
and $TP_{WKB}$ for different values of $E_m$ and $\Delta t$ for the case with $A = 0.001$ 
and $B = 0.001$. We can see, clearly, that both tunneling probabilities coincide 
if we consider the appropriate $\Delta t$, for each $E_m$.

\begin{table}[h!]
{\scriptsize\begin{tabular}{|c|c|c|c|}
\hline $E_m$ & $TP$ & $TP_{WKB}$ & $\Delta t$ \\ \hline
$79$ & $5.74738\times 10^{-303}$ & $8.88570\times 10^{-303}$ & $12.5$\\ \hline
$99$ & $1.22758\times 10^{-256}$ & $3.03907\times 10^{-257}$ & $18.5$\\ \hline
$119$ & $8.00611\times 10^{-212}$ & $2.78301\times 10^{-213}$ & $26$\\ \hline
$139$ & $9.09170\times 10^{-169}$ & $1.31219\times 10^{-170}$ & $33$\\ \hline
$159$ & $7.49408\times 10^{-129}$ & $5.04561\times 10^{-129}$ & $45$\\ \hline
$179$ & $1.78402\times 10^{-89}$ & $2.22444\times 10^{-88}$ & $57$\\ \hline
$199$ & $3.42155\times 10^{-48}$ & $1.46158\times 10^{-48}$ & $73$\\ \hline
$209$ & $6.57786\times 10^{-29}$ & $6.20233\times 10^{-29}$ & $82$\\ \hline
$216$ & $8.14639\times 10^{-16}$ & $2.67764\times 10^{-15}$ & $89$\\ \hline
$222$ & $5.65591\times 10^{-03}$ & $1.11987\times 10^{-03}$ & $100$\\ \hline
\end{tabular}
}
\caption{{\protect\footnotesize {A comparison between $TP$ and $TP_{WKB}$
for $10$ different values of $E_m$ with its associated integration time $\Delta t$ 
for the case with $A = 0.001$ and $B = 0.001$.
}}}
\label{comparison}
\end{table}

In order to have an idea on how $TP$ may differ from the $TP_{WKB}$,
we let the initial wave-packet (\ref{sec5eq1}), with different mean energies, 
evolve during the same time interval $\Delta t$. We consider the example given 
above, in this section, with a common time interval of 100.
We show this comparison in Table \ref{tabela1}, in the appendix, where we have 
an entry for $TP_{WKB}$. It means that, we computed the values of the $TP_{WKB}$s 
for each energy used to compute the $TP$s, in the case where $A = 0.001$ and 
$B = 0.001$. As we can see from Table \ref{tabela1}, for this choice of $\Delta t$ 
the tunneling probabilities disagree for most values of $E_m$. They only agree for
values of $E_m$ very close to the top of the potential barrier. There, because
the values of $E_m$ are similar to $222$, $\Delta t$ is almost sufficient to 
guarantee that $\Psi_R$ of each wave-packet does not contribute to the 
$TP$.

\noindent

\subsection{Tunneling probability as a function of $A$ and $B$}
\label{subsec:tpa}

The Chaplygin gas has two parameters $A$ and $B$ equation (\ref{sec1eq2}). In the present
subsection, we study the behavior of $TP$ equation (\ref{sec5eq1}) as a function of both
parameters. It is important to notice from the Chaplygin potential, equation (\ref{sec1eq2}),
that the parameter $A$ behaves as $a$ increases very much like a positive cosmological constant.
Therefore, we expect that $TP$ grows with $A$ in the same way as it grows with the cosmological
constant \cite{gil}. On the other hand, the dependence of $TP$ on the parameter $B$ can not
be inferred from previous works.

\begin{figure}[h!]
\includegraphics[width=8cm,height=12cm, angle=-90]{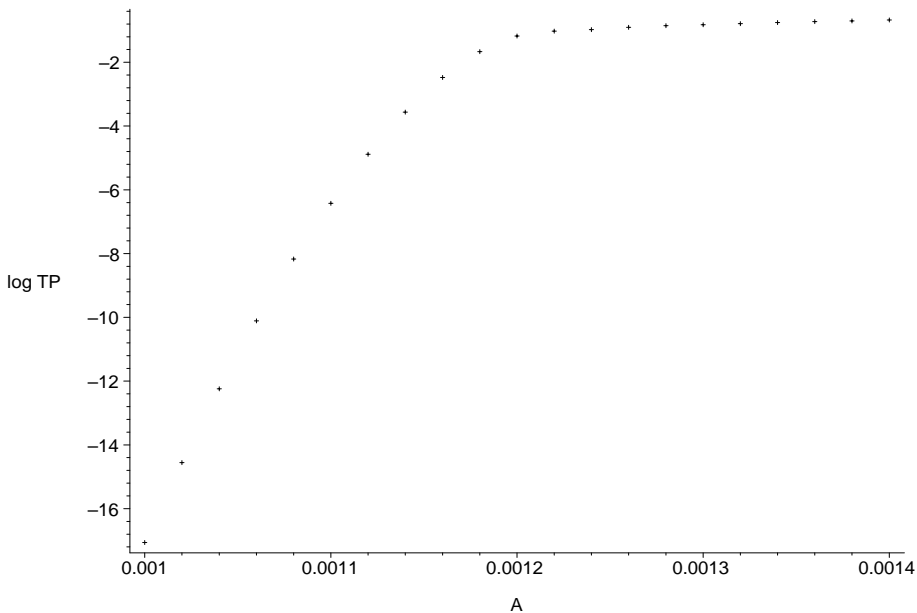}
\caption{Tunneling probability as a function of the parameter $A$ for $E_m=187$ and 
$B=0.001$.}
\label{chaplyginfigure5}
\end{figure}

We start studying the dependence of $TP$ on the parameter $A$. In 
order to do that, we must fix the initial energy $E_m$ for the radiation 
and the parameter $B$. Then, we must compute the $TP$ for various values 
of the parameter $A$. From that numerical study we conclude that the 
tunneling probability grows with $A$ for fixed $E_m$ and $B$. As an 
example, we consider $21$ values of $A$ for fixed $E_m=187$ and $B=0.001$, 
such that, the maximum energy of the potential barrier 
($PE_{max}$), for each $A$, is greater than $187$. The values of $A$, 
$TP$, $a_1$ and $a_2$ are given in table \ref{tableA}, 
in the appendix. With those values, we construct the curve 
$TP$ versus $A$, shown in figure \ref{chaplyginfigure5}. Due to the small 
values of some $TP'$s, we plot the logarithms of the $TP'$s against $A$. 
From figure \ref{chaplyginfigure5}, it is clear that the tunneling 
probability increases with $A$ for a fixed $E_m$ and $B$, as expected. 
Therefore, it is more likely for the universe, described by the present 
model, to nucleate with the highest possible value of the parameter $A$.
That result is in agreement with the approximated, analytical expression 
for $TP$ as a function of $A$, obtained in Reference \cite{moniz}, with the 
boundary conditions of the tunneling wave-function \cite{vilenkin}.


\begin{figure}[h!]
\includegraphics[width=8cm,height=12cm, angle=-90]{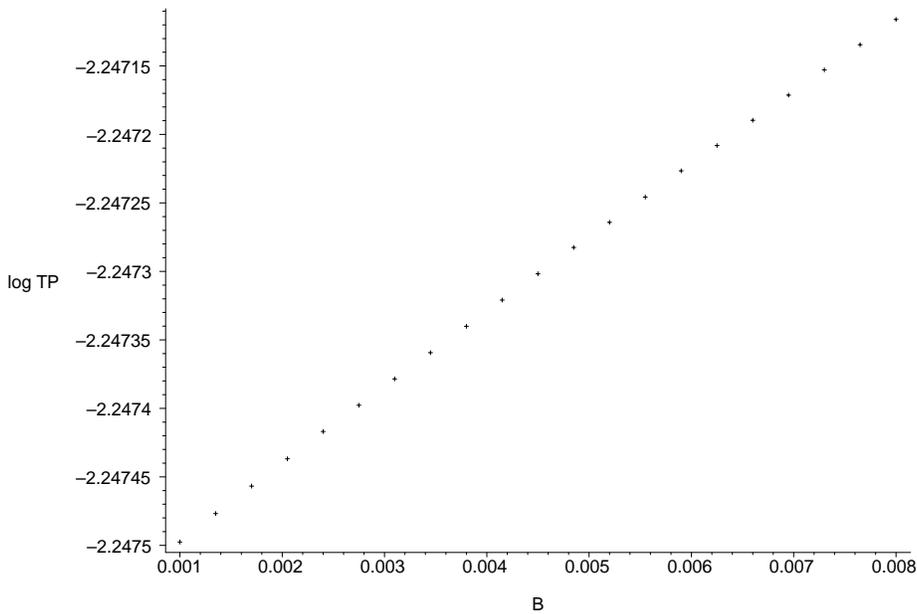}
\caption{Tunneling probability as a function of the parameter $B$ for $E_m=222$ and 
$A=0.001$.}
\label{chaplyginfigure6}
\end{figure}

Let us consider, now, the dependence of $TP$ on the parameter $B$. In 
order to do that, we must fix the initial energy $E_m$ for the radiation 
and the parameter $A$. Then, we must compute the $TP$ for various values 
of the parameter $B$. From that numerical study we conclude that the 
tunneling probability grows with $B$ for fixed $E_m$ and $A$. As an 
example, we consider $21$ values of $B$ for fixed $E_m=222$ and $B=0.001$, 
such that, the maximum energy of the potential barrier 
($PE_{max}$), for each $B$, is greater than $222$. The values of $B$, 
$TP$, $a_1$ and $a_2$ are given in table \ref{tableB}, 
in the appendix. With those values, we construct the curve 
$TP$ versus $B$, shown in figure \ref{chaplyginfigure6}. Due to the small 
values of some $TP'$s, we plot the logarithms of the $TP'$s against $B$. 
From figure \ref{chaplyginfigure6}, it is clear that the tunneling 
probability increases with $B$ for a fixed $E_m$ and $A$. 
Therefore, it is more likely for the universe, described by the present 
model, to nucleate with the highest possible value of the parameter $B$.
That result is in agreement with the approximated, analytical expression 
for $TP$ as a function of $B$, obtained in Reference \cite{moniz}, with the 
boundary conditions of the tunneling wave-function \cite{vilenkin}.

\section{Initial conditions for inflation}

Based on these results one can evaluate the duration of the Planck era according to 
these models. In other words, the time that the universe takes to emerge at the right 
side of the potential barrier. According to the literature \cite{griffths}, the time
that a particle takes to cross a potential barrier is given by
\begin{equation}
t_{planck} = \frac{2a_{1}}{TP},
\label{sec5eq3}
\end{equation}

\noindent
where $a_{1}$ corresponds to the turning point to the left of the barrier and $TP$ is 
the tunneling probability Eq. (\ref{sec5eq1}). We have computed $t_{planck}$, for all
examples considered here. In Tables \ref{tabela1}, \ref{tableA} and \ref{tableB},
in the appendix, we have an entry to $t_{planck}$. 

After the time interval $t_{planck}$ the universe emerges to the right of the
barrier with $a_2$ and $p_a = 0$, because $a_2$ is a turning point. Now, one may use
these values as initial conditions to compute the dynamical evolution of the scalar
factor through the integration of the appropriate classical Hamilton's equations. Due
to our potential the classical evolution of $a$ will correspond to an inflationary phase 
of the model, posterior to the Planck era. In the inflationary phase, according to the 
literature \cite{guth}, the universe has increased its size in many orders of magnitude, 
in a very short time. The table \ref{tabela1}, in the appendix, shows the initial conditions 
($a_2$) for different values of $E_{m}$, where the corresponding moments $p_{a}$ always 
vanish. Applying the 8th order Runge-Kuta method, implemented through a numerical routine in 
MAPLE, we have integrated the Hamilton's equations for many sets of initial conditions, in 
order to calculate the time evolution of the scale factor (conformal time). We have concluded
from that numerical investigation that the scale factor always increases many orders of
magnitude in very short period of time, which confirms the development of a inflationary
phase just after the Planck era.

\begin{figure}[h!]
\includegraphics[width=6cm,height=10cm, angle=-90]{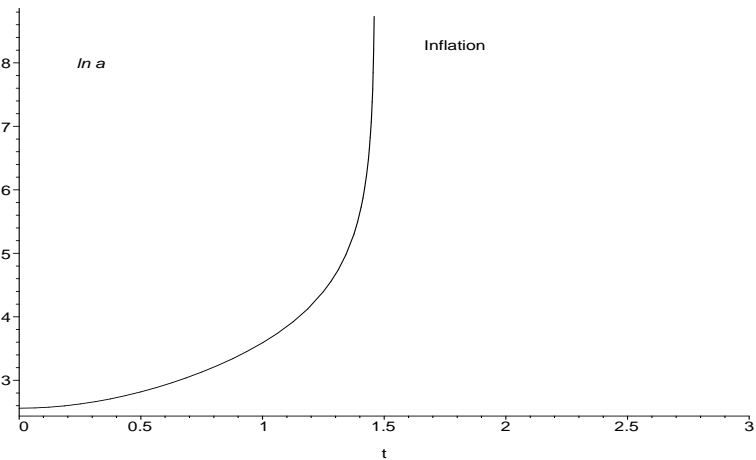}
\vspace{0.8cm}
\caption{Scale factor as a function of the conformal time (logarithmic scale). The initial conditions are $a_{2}=12.95147904$ and $p_{a}=0$ in $\tau = 0$, after the Planck era for wave packets with $E_{m}=220$.}
\label{g1}
%
%
\end{figure}

As an example, consider the case where the wave packet has $E_{m}=220$ and initial conditions 
$a_{2}=12.95147904$ and $p_{a}=0$ at $\tau = 0$, the scale factor increases from $12.95147904$ 
to $11465.3370549124$ in the interval $\Delta \tau =1.46$. This behavior is shown in figure 
\ref{g1}, which presents (in logarithm scale) the time evolution of the scale factor $a(\tau)$, 
as a function of the conformal time.

\section{Conclusion and final remarks}

The quantization of a FRW model with radiation and Chaplygin gas was made via finite
differences method and the Crank-Nicholson scheme. We considered initial wave packets with
finite norm, and well defined values of the mean energy, associated to the radiation fluid, and
the parameters of the Chaplygin gas. Using this numerical method, one builds the time evolution
of these wave packets. The results show that the quantum dynamics of these models may be
described by wave packets with finite norm, which satisfy appropriate boundary conditions. In
other words, such wave packets are well-defined over the whole space, even when the scale
factor goes to zero.

Observing these wave packets, one can see the presence of strong oscillations for the values of the scale factor, which precede the effective potential barrier of the model. Then, when the packets reach the potential barrier, the amplitude of those oscillations decrease drastically, vanishing at a point far away from the barrier, at infinity. This is the quantum tunneling, which explains the birth of the universe with a certain, well-defined, size after the tunneling. Via numerical calculations, it was possible to determine the dependence of the tunneling probability with $E_m$, $A$ and $B$. We also compared it with the tunneling probability obtained using the WKB approximation. The results showed that both increase with the mean energy of the initial wave packets and show a good agreement, under certain circumstances.
We verified that the duration of the quantum phase depends on the mean energy assigned to the initial wave packet, i.e., the bigger the energy, the shorter is the duration of the Planck era. We also, determined the conditions of the universe at the end of the Planck era. These
conditions, obtained by the quantization of the model, were used as initial conditions to determine the classical evolution of the scale factor of the universe. The evolution shows that, after its quantum phase, the universe undergoes a very accelerated growth. It increases its size in many orders of magnitude in a very short time interval. We identify this phase as the inflationary phase.

\begin{acknowledgements}
E. V. Corr\^{e}a Silva (Researcher of CNPq, Brazil), L. G. Ferreira Filho, G. A. Monerat and
G. Oliveira-Neto thank CNPq for partial financial support (Edital Universal CNPq/476852/2006-4). G. A. Monerat thanks FAPERJ for partial financial support (contract E-26/170.762/2004). P. Romildo Jr. thanks CAPES of Brazil, for his scholarship. F.G. Alvarenga, J.C. Fabris and S.V.B. Gon\c{c}alves also thank CNPq for partial financial support. R. Fracalossi thanks CAPES of Brazil for financial support.
\end{acknowledgements}

\appendix

\section{Tables}

\begin{table}[h!]
{\scriptsize\begin{tabular}{|c|c|c|c|c|c|} 
\hline Energy ($E_m$) & PT & $t_{planck}$ & $a_{1}$ & $a_{2}$ & $PT_{WKB}$\\ \hline 
222.000000 & 0.00565591 & 4134.35 & 11.691761 & 12.701983 & $1.1198667720\cdot 10^{-3}$\\ \hline 
221.000000 & 0.000963791 & 23947 & 11.539957 & 12.840057  & $1.307940728\cdot 10^{-5}$\\ \hline 
220.000000 & 0.0001864 & 122476 & 11.414765 & 12.951479 & $1.521014497\cdot 10^{-7} $\\ \hline 
219.000000 & $3.99923\cdot 10^{-5}$ & 565372 & 11.305263 & 13.047172 & $1.762093474\cdot 10^{-9}$\\ \hline 
218.000000 & $9.35514\cdot 10^{-6}$ & $2.39578\cdot 10^{6}$ & 11.206415 & 13.132172 & $2.033617858\cdot 10^{-11}$\\ \hline 
217.000000 & $2.35555\cdot 10^{-6}$ & $9.43763\cdot 10^{6}$ & 11.115403 & 13.209295 & $2.338009646\cdot 10^{-13}$\\ \hline 
216.000000 & $6.32225\cdot 10^{-7}$ & $3.48941\cdot 10^{7}$ & 11.030465 & 13.280306 & $2.677635300\cdot 10^{-15}$\\ \hline 
215.000000 & $1.79509\cdot 10^{-7}$ & $1.22004\cdot 10^{8}$ & 10.950407 & 13.346395 &$3.054754108\cdot 10^{-17}$\\ \hline 
214.000000 & $5.3592\cdot 10^{-8}$ & $4.05821\cdot 10^{8}$ & 10.874382 & 13.408411 & $3.471460808\cdot 10^{-19}$\\ \hline 
213.000000 & $1.674\cdot 10^{-8}$ & $1.29054\cdot 10^{9}$ & 10.801756 & 13.466987 & $3.929622371\cdot 10^{-21}$\\ \hline 
212.000000 & $5.4483\cdot 10^{-9}$ & $3.9396\cdot 10^{9}$ & 10.732048 & 13.522604 & $4.430810920\cdot 10^{-23}$\\ \hline 
211.000000 & $1.84123\cdot 10^{-9}$ & $1.15845\cdot 10^{10}$ & 10.664877 & 13.575643 & $4.976232532\cdot 10^{-25}$\\ \hline 
210.000000 & $6.44182\cdot 10^{-10}$ & $3.29098\cdot 10^{10}$ & 10.599935 & 13.626410 & $5.566654752\cdot 10^{-27}$\\ \hline 
209.000000 & $2.32726\cdot 10^{-10}$ & $9.05525\cdot 10^{10}$ & 10.536974 & 13.675154 & $6.202333780\cdot 10^{-29}$\\ \hline 
208.000000 & $8.66257\cdot 10^{-11}$ & $2.41863\cdot 10^{11}$ & 10.475786 & 13.722084 & $6.882942312\cdot 10^{-31}$\\ \hline 
207.000000 & $3.31557\cdot 10^{-11}$ & $6.28321\cdot 10^{11}$ & 10.416196 & 13.767372 & $7.607503240\cdot 10^{-33}$\\ \hline 
206.000000 & $1.30263\cdot 10^{-11}$ & $1.59033\cdot 10^{12}$ & 10.358055 & 13.811169 & $8.374327820\cdot 10^{-35}$\\ \hline 
205.000000 & $5.24514\cdot 10^{-12}$ & $3.92792\cdot 10^{12}$ & 10.301236 & 13.853600 & $9.18096264\cdot 10^{-37}$\\ \hline 
204.000000 & $2.16153\cdot 10^{-12}$ & $9.47999\cdot 10^{12}$ & 10.245627 & 13.894776 & $1.002414945\cdot 10^{-38}$\\ \hline 
203.000000 & $9.10506\cdot 10^{-13}$ & $2.23856\cdot 10^{13}$ & 10.191134 & 13.934793 & $1.089979620\cdot 10^{-40}$\\ \hline 
202.000000 & $3.91585\cdot 10^{-13}$ & $5.17776\cdot 10^{13}$ & 10.137670 & 13.973737 & $1.180296533\cdot 10^{-42}$\\ \hline 
201.000000 & $1.71766\cdot 10^{-13}$ & $1.17429\cdot 10^{14}$ & 10.085160 & 14.011682 & $1.272788104\cdot 10^{-44}$\\ \hline 
200.000000 & $7.67719\cdot 10^{-14}$ & $2.61386\cdot 10^{14}$ & 10.033537 & 14.048694 & $2.059920186\cdot 10^{-46}$\\ \hline 
199.000000 & $3.49333\cdot 10^{-14}$ & $5.71532\cdot 10^{14}$ & 9.982742 & 14.084834  & $1.461582575\cdot 10^{-48}$\\ \hline 
189.000000 & $3.05503\cdot 10^{-17}$ & $6.22616\cdot 10^{17}$ & 9.510554 & 14.407881 & $2.255557889\cdot 10^{-68}$\\ \hline 
179.000000 & $8.61288\cdot 10^{-20}$ & $2.10926\cdot 10^{20}$ & 9.083391 & 14.680929 & $2.224440858\cdot 10^{-88}$\\ \hline 
169.000000 & $5.67252\cdot 10^{-22}$ & $3.06189\cdot 10^{22}$ & 8.684327 & 14.920460 & $1.364184268\cdot 10^{-108}$\\ \hline 
159.000000 & $7.12611\cdot 10^{-24}$ & $2.33052\cdot 10^{24}$ & 8.303768 & 15.135558 & $5.045605144\cdot 10^{-129}$\\ \hline 
149.000000 & $1.48584\cdot 10^{-25}$ & $1.06815\cdot 10^{26}$ & 7.935486 & 15.331854 & $1.087228346\cdot 10^{-149}$\\ \hline 
139.000000 & $4.64724\cdot 10^{-27}$ & $3.26\cdot 10^{27}$ & 7.575013 & 15.513119 & $1.312194871\cdot 10^{-170}$\\ \hline 
129.000000 & $2.01877\cdot 10^{-28}$ & $7.15172\cdot 10^{28}$ & 7.218847 & 15.682026 & $8.478448344\cdot 10^{-192}$\\ \hline 
119.000000 & $1.14576\cdot 10^{-29}$ & $1.19816\cdot 10^{30}$ & 6.864019 & 15.840547 & $2.783009253\cdot 10^{-213}$\\ \hline 
109.000000 & $8.07577\cdot 10^{-31}$ & $1.61169\cdot 10^{31}$ & 6.507811 & 15.990187 & $4.364145700\cdot 10^{-235}$\\ \hline 
99.000000 & $6.76014\cdot 10^{-32}$ & $1.81876\cdot 10^{32}$ & 6.147539 & 16.132125 & $3.039077147\cdot 10^{-257}$\\ \hline 
89.000000 & $6.44132\cdot 10^{-33}$ & $1.79477\cdot 10^{33}$ & 5.780359 & 16.267303 & $8.604523368\cdot 10^{-280}$\\ \hline 
79.000000 & $6.68111\cdot 10^{-34}$ & $1.61741\cdot 10^{34}$ & 5.403039 & 16.396490 & $8.885695860\cdot 10^{-303}$\\ \hline 
69.000000 & $7.14821\cdot 10^{-35}$ & $1.40222\cdot 10^{35}$ & 5.011661 & 16.520320 & $2.918766558\cdot 10^{-326}$\\ \hline 
59.000000 & $7.31694\cdot 10^{-36}$ & $1.25767\cdot 10^{36}$ & 4.601147 & 16.639326 & $2.553230826\cdot 10^{-350}$\\ \hline 
49.000000 & $6.33942\cdot 10^{-37}$ & $1.31383\cdot 10^{37}$ & 4.164443 & 16.753960 & $4.679948348\cdot 10^{-375}$\\ \hline 
39.000000 & $3.66812\cdot 10^{-38}$ & $2.01243\cdot 10^{38}$ & 3.690908 & 16.864607 & $1.278053730\cdot 10^{-400}$\\ \hline 
29.000000 & $8.13428\cdot 10^{-40}$ & $7.77617\cdot 10^{39}$ & 3.162678 & 16.971602 & $1.801636092\cdot 10^{215}$\\ \hline 
19.000000 & $1.21515\cdot 10^{-42}$ & $4.18787\cdot 10^{42}$ & 2.544451 & 17.075236 & $2.645204325\cdot 10^{-455}$\\ \hline 
9.000000 & $5.24188\cdot 10^{-51}$ & $6.64299\cdot 10^{50}$ & 1.741086 & 17.175763 & $1.109066979\cdot 10^{-485}$\\ \hline 
8.000000 & $1.28841\cdot 10^{-51}$ & $2.54671\cdot 10^{51}$ & 1.640607 & 17.185653 & $6.395403152\cdot 10^{-489}$\\ \hline 
7.000000 & $8.60797\cdot 10^{-52}$ & $3.56372\cdot 10^{51}$ & 1.533820 & 17.195515 & $3.262847086\cdot 10^{-492}$\\ \hline 
6.000000 & $4.88772\cdot 10^{-52}$ & $5.80764\cdot 10^{51}$ & 1.419304 & 17.205349 & $1.446721601\cdot 10^{-495}$
\\ \hline 
5.000000 & $3.13507\cdot 10^{-52}$ & $8.26144\cdot 10^{51}$ & 1.295012 & 17.215155 & $5.443204796\cdot 10^{-499}$\\ \hline 
4.000000 & $1.78021\cdot 10^{-52}$ & $1.30074\cdot 10^{52}$ & 1.157802 & 17.224932 & $1.680368783\cdot 10^{-502}$\\ \hline 
3.000000 & $9.70346\cdot 10^{-53}$ & $2.06603\cdot 10^{52}$ & 1.002384 & 17.234682 & $4.045122584\cdot 10^{-506}$\\ \hline 
2.000000 & $4.42249\cdot 10^{-53}$ & $3.70114\cdot 10^{52}$ & 0.818412 & 17.244404 &$6.963887216\cdot 10^{-510}$\\ \hline 
1.000000 & $1.31081\cdot 10^{-53}$ & $8.8352\cdot 10^{+52}$ & 0.579062 & 17.254099 & $7.140705628\cdot 10^{-514}$\\ \hline 
\end{tabular}}
\caption{{\protect\footnotesize {Relation between the mean energies of the wave packets and the tunneling probability for $A=0.001$; $B=0.001$; $N=4000$ (spacial discretization); $a_{max}=30$; $tmax=100$; and $V_{max}=223.5282023$, which corresponds to the maximum value of the potential. Here, we see as well the initial conditions $a_2$, for the Universe to emerge from the Planck era to the Inflationary phase, all of them with $p_{a}=0$.}}}
\label{tabela1}
\end{table}

\begin{table}[h!]
{\scriptsize\begin{tabular}{|c|c|c|c|c|} 
\hline $A$ & $TP$ & $t_{planck}$ & $a_1$ & $a_2$ \\ \hline 
0.001000 & 8.70789e-18 & 2.16406e+18 & 9.422222 & 14.465801 \\ \hline 
0.001020 & 2.77614e-15 & 6.81336e+15 & 9.457412 & 14.340803 \\ \hline 
0.001040 & 5.7598e-13 & 3.2965e+13 & 9.493610 & 14.216939 \\ \hline 
0.001060 & 7.76552e-11 & 2.45467e+11 & 9.530909 & 14.094026 \\ \hline 
0.001080 & 6.76155e-09 & 2.83054e+09 & 9.569412 & 13.971873 \\ \hline 
0.001100 & 3.75605e-07 & 5.11668e+07 & 9.609242 & 13.850280 \\ \hline 
0.001120 & 1.30432e-05 & 1.47978e+06 & 9.650536 & 13.729031 \\ \hline 
0.001140 & 0.00027424 & 70693.3 & 9.693459 & 13.607891 \\ \hline 
0.001160 & 0.00332339 & 5860.41 & 9.738201 & 13.486603 \\ \hline 
0.001180 & 0.0215238 & 909.224 & 9.784991 & 13.364874 \\ \hline 
0.001200 & 0.0668755 & 294.102 & 9.834106 & 13.242367 \\ \hline 
0.001220 & 0.0946396 & 208.916 & 9.885885 & 13.118686 \\ \hline 
0.001240 & 0.105477 & 188.492 & 9.940751 & 12.993352 \\ \hline 
0.001260 & 0.125288 & 159.62 & 9.999247 & 12.865774 \\ \hline 
0.001280 & 0.139881 & 143.866 & 10.062084 & 12.735189 \\ \hline 
0.001300 & 0.149677 & 135.361 & 10.130225 & 12.600590 \\ \hline 
0.001320 & 0.162884 & 125.305 & 10.205035 & 12.460569 \\ \hline 
0.001340 & 0.175986 & 116.925 & 10.288549 & 12.313046 \\ \hline 
0.001360 & 0.187801 & 110.586 & 10.384061 & 12.154690 \\ \hline 
0.001380 & 0.19885 & 105.583 & 10.497560 & 11.979472 \\ \hline 
0.001400 & 0.212192 & 100.309 & 10.642415 & 11.773989 \\ \hline 
\end{tabular}}
\caption{{\protect\footnotesize {The computed values of $TP$, $A$,
$t_{planck}$, $a_1$ and $a_2$ for $21$ different values of $A$
when $E=187$ and $B=0.001$.}}}
\label{tableA}
\end{table}

\begin{table}[h!]
{\scriptsize\begin{tabular}{|c|c|c|c|c|} 
\hline $B$ & $TP$ & $t_{planck}$ & $a_1$ & $a_2$ \\ \hline 
0.001000 & 0.00565591 & 4134.35 & 11.691761 & 12.701983 \\ \hline 
0.001350 & 0.00565618 & 4134.16 & 11.691764 & 12.701982 \\ \hline 
0.001700 & 0.00565644 & 4133.97 & 11.691766 & 12.701980 \\ \hline 
0.002050 & 0.0056567 & 4133.78 & 11.691768 & 12.701978 \\ \hline 
0.002400 & 0.00565696 & 4133.59 & 11.691770 & 12.701977 \\ \hline 
0.002750 & 0.00565721 & 4133.4 & 11.691773 & 12.701975 \\ \hline 
0.003100 & 0.00565746 & 4133.22 & 11.691775 & 12.701973 \\ \hline 
0.003450 & 0.00565771 & 4133.04 & 11.691777 & 12.701971 \\ \hline 
0.003800 & 0.00565796 & 4132.86 & 11.691779 & 12.701970 \\ \hline 
0.004150 & 0.00565821 & 4132.68 & 11.691781 & 12.701968 \\ \hline 
0.004500 & 0.00565846 & 4132.5 & 11.691784 & 12.701966 \\ \hline 
0.004850 & 0.00565871 & 4132.32 & 11.691786 & 12.701964 \\ \hline 
0.005200 & 0.00565895 & 4132.14 & 11.691788 & 12.701963 \\ \hline 
0.005550 & 0.00565919 & 4131.96 & 11.691790 & 12.701961 \\ \hline 
0.005900 & 0.00565944 & 4131.79 & 11.691793 & 12.701959 \\ \hline 
0.006250 & 0.00565968 & 4131.61 & 11.691795 & 12.701958 \\ \hline 
0.006600 & 0.00565992 & 4131.43 & 11.691797 & 12.701956 \\ \hline 
0.006950 & 0.00566016 & 4131.26 & 11.691799 & 12.701954 \\ \hline 
0.007300 & 0.0056604 & 4131.08 & 11.691801 & 12.701952 \\ \hline 
0.007650 & 0.00566064 & 4130.91 & 11.691804 & 12.701951 \\ \hline 
0.008000 & 0.00566088 & 4130.74 & 11.691806 & 12.701949 \\ \hline 
\end{tabular}}
\caption{{\protect\footnotesize {The computed values of $TP$, $B$,
$t_{planck}$, $a_1$ and $a_2$ for $21$ different values of $B$
when $E=222$ and $A=0.001$.}}}
\label{tableB}
\end{table}
\end{document}